# Periodic negative differential conductance in a single metallic nano-cage

Yehonadav Bekenstein[1,2], Kathy Vinokurov[2], Tal J. Levy[3], Eran Rabani[3*], Uri Banin[2*], Oded Millo[1*]

[1]*Racah Institute of Physics and the Center for Nanoscience and Nanotechnology, The Hebrew University of Jerusalem, Jerusalem 91904, Israel*

[2] *Institutes of Chemistry and the Center for Nanoscience and Nanotechnology, The Hebrew University of Jerusalem, Jerusalem 91904, Israel*

[3] *School of Chemistry, Sackler Faculty of Exact Sciences, Tel Aviv University, Tel Aviv 69978, Israel*

*To whom correspondence should be addressed. e-mail: milode@vms.huji.ac.il , uri.banin@huji.ac.il, rabani@tau.ac.il

## ABSTRACT

We report a bi-polar multiple periodic negative differential conductance (NDC) effect on a single cage-shaped Ru nanoparticle measured using scanning tunneling spectroscopy. This phenomenon is assigned to the unique multiply-connected cage architecture providing two (or more) defined routes for charge flow through the cage. This, in turn, promotes a self-gating effect, where electron charging of one route affects charge transport along a neighboring channel, yielding a series of periodic NDC peaks. This picture is established and analyzed here by a theoretical model.

Multiple negative differential conductance (NDC) [1-3], in which increase in voltage leads to decrease in current at several consecutive voltage values, has been reported previously for resonant tunneling-diode devices based on semiconductor heterostructures [4-5]. The multiple NDC effect was utilized in functional electronic devices to realize multiple value logic, ultra-high speed analog-to-digital converters, frequency multipliers and other circuit elements[6-7]. However, these devices require relatively complex fabrication with characteristic micron-scale dimensions, and typically exhibit only 2-4 non-periodic NDC peaks for one bias polarity. It is of interest to reduce the device size, on one hand, and obtain a bipolar multiple periodic NDC effect, on the other hand. This was discovered by us for single empty Ru nano-cages, where the tunneling I-V (current-voltage) characteristics measured using scanning tunneling spectroscopy (STS) exhibit up to six periodic NDC peaks, for both bias polarities. As shown below, quite frequently the expected 'conventional' Coulomb staircase effect [8] measured on a Ru nano-cage surprisingly evolved into a series of NDC peaks, while nearly maintaining the staircase periodicity (as a function of bias). As demonstrated by a model simulation, this intriguing phenomenon is well accounted for by the unique multiply-connected cage architecture, which enables a self-gating like effect between neighboring transport channels through the different cage arms.

We study here Ru nano-cages, whose discovery was reported by us recently [9-10]. These are synthesized starting from a hybrid semiconductor/metal quantum-dot (QD) comprising a metallic Ru cage-like shell grown selectively on the edges of a semiconducting $Cu_2S$ nanocrystal. The empty Ru cages were then obtained by methanol addition to a solution of $Cu_2S$/Ru hybrids in toluene, leading to selective dissolution of the $Cu_2S$ from the cage interior.

To study the electronic and transport properties of this system, we utilized scanning tunneling microscopy and spectroscopy (STM & STS). These are effective tools for studying the electrical properties of nano-structured systems. They are particularly suitable for hybrid and cage-like QDs due to the ability of measuring the local density of states (DOS) with nanometric spatial resolution [11-14]. For the STM measurements, QDs solutions were drop cast onto a flame annealed Au(111) substrate and let dry [Fig. S1 in the supporting information (SI)]. The STM measurements were performed at 4.2 K, using Pt-Ir tips, in clean He exchange gas inserted into the sample space after evacuation. Tunneling current-voltage (I-V) characteristics were acquired after positioning the STM tip at different locations above individual QDs, realizing a double barrier tunnel junction (DBTJ) configuration [8], and disabling momentarily the feedback loop. The dI/dV-V tunneling spectra, were numerically derived from the measured I-V curves. The topographic images were acquired with current and sample-bias set values of $I_s \cong 0.1$ nA and $V_s \cong 1$ V. 20 Ru cages were measured, out of which 8 showed periodic NDC effect. No such effect was found on any of the 10 $Cu_2S$/Ru hybrids that were measured, which exhibited only the conventional single electron tunneling (SET) behavior (on the Ru cage).

STM images of single Ru cages having two different orientations deposited on a conducting Au(111) substrate (Fig. S1) along with the corresponding STEM (scanning transmission electron microscope) images and illustrations are presented in Figs. 1(a-f). The two different projections demonstrate the cage structure, depicting the median arm and the pore of the cage. Figure 1 also presents I-V characteristics and the corresponding dI/dV-V spectra acquired on the Ru cage (inset). The blue curves exhibit conventional SET effects [15-16], the Coulomb blockade and staircase, commonly observed in tunneling through metallic nanostructures [8]. The former manifests itself in the suppression of the tunneling current and the DOS around zero bias, while the latter by a periodic series of broadened steps (peaks) in the I-V (dI/dV-V) characteristics, each corresponding to the addition of a single electron to the Ru-cage. Surprisingly, however, in many cases we observed a set of periodic

NDC peaks, as shown by the green curves. We have verified that the oscillation period, as a function of bias, was independent of the bias sweep-rate over a very wide range, 10 to 500 V/s, thus ruling out the possibility that these features are associated with pick-up of external noise. The I-V curves on a given cage showed, in different scans, either the 'conventional' Coulomb blockade and staircase behavior (blue curves), or the periodic NDC effect (green curves), while roughly maintaining the same periodicity, pointing out to a connection between the two phenomena. The transition between the staircase and NDC behaviors was accompanied by a change (increase or decrease) of the overall tunneling resistance.

The inset of Fig. 1(g) depicts the bias values of the peaks of the dI/dV-V curves for both NDC and SET (staircase) data [seen in Figs. 1(g) and 1(h)] as a function of peak number (negative for negative bias values). In the SET case it is well established that each peak is due to a change by one in the number of excess electrons on the cage, and the average single electron charging energy can be readily extracted from this plot, $U \sim 130\ meV$. Using the simplistic formula $U = e^2/2C$, an average effective capacitance of $C = 6.2 \cdot 10^{-19} F$ is obtained for this tip-QD-substrate configuration, while fit to the 'orthodox model' [8] for SET yields comparable values for the tip-cage and cage-substrate junction capacitances, $C_1$ = $4 \cdot 10^{-19} F$ and $C_2$ = $13 \cdot 10^{-19} F$, respectively (Fig. S11) Remarkably, the peak spacing in the NDC case nearly coincides with that of the staircase, indicating that the underlying NDC mechanism must be associated with single electron charging of the cage. Note, however, that there is a small shift between the two sets of peaks, a point that is addressed below. The single electron charging energies measured on all the empty Ru nano-cages largely varied from one QD to another, between ~100 to ~300 meV, as shown by Fig. S12, yet in all cases where NDC peaks emerged, their periodicity well corresponded to the staircase periodicity, as in the cases depicted by Figs. 1 and S13. This large spread in charging energies may be attributed mainly and to variations in the tunnel junction parameters between measurements, and also to the spread in the widths of the arms between nano-cages, but not solely to variations in their diameters that have a rather narrow size distribution of less than 13% (Fig. S12).

NDC effects have been observed previously in resonant tunneling diodes made of micron-size semiconductor heterostrucutres [2-3] and on various DBTJ configurations, either lithographically-defined in two-dimensional electron-gas [17] or achieved, as in our case here, in STS measurements of QDs [11, 18]. However, bi-polar periodic NDC oscillations correlated with the Coulomb staircase, as well as the bi-stability between SET and NDC behavior, have not yet been reported. The origin of the periodic NDC is attributed to the special multiply-connected geometry of the empty cage QD, providing multiple defined routes for charge flow through the cage. This, in turn, may promote a self-gating like effect, where electron charging of one route may influence current flow through neighboring routes [19-20], yielding a series of NDC peaks with the Coulomb staircase periodicity as a function of bias voltage.

To study this conjecture we devised a simplified model [16, 21] that considers two coupled conducting channels provided by the Ru cage, each supporting several charging levels. The two channels are also connected by tunneling barriers (i.e., weakly coupled) to the STM tip and the conducting substrate. A model Hamiltonian describing the electronic structure of the Ru cage is given by:

$$H_{cage} = \varepsilon \sum_{\nu=\alpha,\beta} \sum_i n_i^\nu + U \sum_{\nu=\alpha,\beta} \sum_{i>j} n_i^\nu n_j^\nu + U_{int} \sum_{i,j} n_i^\alpha n_j^\beta .$$

The first term in $H_{cage}$ represents the non-interacting energy for each state $i$ on the two channels, $\nu = \alpha, \beta$, which also models the effects of the Fermi level offset affecting the Coulomb blockade. The second term represents the sum of charging energies ($U$) of the two individual channels, where $n_i^\nu = 1\ or\ 0$ if state $i$ on channel $\nu$ is occupied by an electron or not, respectively. This term provides the Coulomb staircase in the case of uncoupled channels. The third term represents the interaction energy ($U_{int}$) between the two channels on the cage.

Figures 2(a) and 2(b) show simulated I-V and dI/dV-V curves calculated using a master equation approach [22] for the above model (see SI for more details). First, we consider the case where the two coupled channels are both connected to the substrate and STM tip with similar tunneling resistances. In this case (blue curves), we observe a periodic Coulomb staircase with spacing $\sim U$ (for simplicity we take $U$ in units of Volts). This equal spacing arises because of the identical charging energies assigned to both channels in the model and choosing $U_{int}$ to be comparable to $U$, while non-equidistant Coulomb steps are obtained when the charging energies differ significantly (not shown). Importantly, NDC is not observed, irrespective of the relative magnitudes of $U$ and $U_{int}$ (SI Fig. S3). Therefore, coupling between the channels by itself does not lead to NDC.

NDC emerges when one of the channels is effectively disconnected from either the STM tip or the substrate (but not from both) and $U_{int}$ is sufficiently large ($U_{int} > 0.6U$) to reduce the current through the conducting channel (SI Fig. S4). Such large values of $U_{int}$ is expected, based on classical electrostatics, when the separation between the arms is smaller than *5R*, where *R* is the radius of the arms (see SI Fig. S9), a condition that is satisfied in our cages, at least near junctions between neighboring arms. $I$-$V$ and $dI/dV$-$V$ characteristics for this case are shown in Figs. 2(a) and 2(b), respectively (green curves). The observed NDC, which is consistent with the experiments, can be traced to a local gating-like effect of the blocked channel on the conducting one. While practically only the fully connected channel contributes to the current, both channels can be charged, as portrayed by the charging level occupation diagrams in Figs. 2(d) and 2(e) for the blocked and conducting channels, respectively. When the bias voltage increases towards $U$, both channels are partially charged and the current increases. Increasing the bias voltage above $U$ leads initially to a decrease in the total current since the charge on the blocked channel increases, hampering conduction through the open channel due to the Coulomb repulsion. Further increase of the bias voltage opens higher charging levels for conduction, and the current increases again while the second charging level of both channels start to populate, reaching a maximum at $V \sim 2U$. Above this bias, the blocked channel is further charged, decreasing again the current in the conducting channel via Coulomb repulsion. The above NDC mechanism repeats itself at bias values with periodicity $U$, with the periodicity arising from consecutive single electron (or hole) charging events in the coupled channels. This mechanism is similar in spirit to NDC induced by populations switching in SET of coupled QDs [11, 17, 23], where the charging level of the blocked channel increases (Fig. 2d) at the account of that of the conducing channel (Fig. 2e). However, while in previous studies only a single event of population switching was observed, the present case gives rise to multiple periodic events, taking place both at positive and negative bias values.

This model captures the essence of transport through the unique multiply connected cage structure and indeed accounts for the observed experimental data. First, the experimental curves showing the NDC (green curves, Fig. 1) have approximately the Coulomb staircase periodicity (blue curves), fully consistent with the model (SI section II for more details). The high symmetry of the cage structure can indeed lead to appearance of two parallel channels with similar charging energies as assumed in the model. Furthermore, the different pathways of conductance in the cage may be disconnected from one another while they are still coupled electrostatically. Indeed, structural analysis with STEM and TEM tomography shows disconnections in the polycrystalline Ru cages [9]. The transition between conventional SET to NDC in the experiments may be attributed to reorientation of the cage or to slight movement of the tip along the nano-cage. This, in turn, can lead to the situation where tunneling resistance in the junction between one of the channels to one lead, most likely the tip, becomes much larger than all others, leading to effective blocking of the current through that channel.

A more subtle observation, that nevertheless requires attention, is the small shift between the peaks of the NDC and Coulomb staircase spectra, seen in both experimental and theoretical curves. The shifts result mainly from changes in the junction parameters taking place due to the aforementioned tip movement and/or QD reorientation that lead to the switching between the conventional SET and NDC behaviors. These, in turn, yield small changes in the charging energy and 'effective residual charge' $Q_0$ [8], as well as in the coupling strengths to the external electrodes. The theoretical curve presented in Fig. 2 was calculated considering only the latter effect, giving rise to an opposite shift compared to that in the experimental curve shown in Fig. 1. Additional experimental spectra, showing other (either positive or negative) shifts between the NDC and SET peaks, along with corresponding theoretical curves, are presented in Fig. S13.

Further insight and support for the periodic NDC mechanism suggested above is provided by the STS results measured on the hybrid $Cu_2S$/Ru-cage QD (before leaching out the $Cu_2S$ core). When positioning the tip on the surrounding Ru cage, tunneling spectra manifesting SET effects were observed, as shown by Fig. 3 for two different bias and current settings. Changing the STM settings affects the tip-QD distance and the corresponding junction capacitance [24], consequently generating a gating-like effect by changing $Q_0$ [8, 25] (and SI section IV). Indeed, the zero-bias gap in the tunneling spectrum presented by the green curves vanished upon modifying the STM setting, taken over by a linear I-V behavior at low bias (Fig. 3(a), blue curves). This behavior, a hallmark of SET effect, establishes that the gap in the former spectrum is due to the Coulomb blockade [8]. At higher bias, both spectra show a Coulomb staircase behavior with similar period of ~315 meV (inset of Fig. 3). This is the largest charging energy observed for the hybrids, whereas the lowest observed value was ~140 meV (Fig. S12).We note in passing that the STS data on the hybrids taken at other positions showed different behaviors, manifesting significantly larger gaps of up to 1.4 eV, corresponding to the band-gap of the semiconducting $Cu_2S$ core decorated by in-gap states.(discussed elsewhere) Importantly, NDC effects were not observed for the hybrid QD irrespective of the STM settings. This is consistent with the hybrid structure that supports and reinforces the external metallic cage structure grown onto it, thus minimizing the possibility of disconnected routes. Additionally, the $Cu_2S$ core can electrically bypass possible disconnections in the Ru cage and also reduce $U_{int}$ by screening the electrostatic coupling between the channels (SI Fig. S6).

In summary, we observed a bi-polar multiple periodic NDC effect on a single cage-shaped Ru nanoparticle using scanning tunneling spectroscopy. A simple model was developed, relating this effect to the unique multiply-connected cage architecture which promotes a self-gating effect, where electron charging of one disconnected route has an effect on a neighboring channel, yielding a series of periodic NDC peaks. This mechanism is expected to be generic and applicable to other types of nanoscale multiply-connected systems that can thus be designed to exhibit multiple NDC with requested periodicity, opening a gateway for the construction of nanoscale electronic devices utilizing this unique effect.

The research received funding from the ERC, FP7 grant agreement n° [246841]. (U.B.), the ISF (O.M.), the US–Israel BSF (E.R.), and the FP7 Marie Curie IOF project  HJSC (E.R) T.J.L. thanks the Nano-Center at Tel Aviv University for a doctoral fellowship. We are grateful to Eran Socher and Avraham Schiller for helpful discussions.

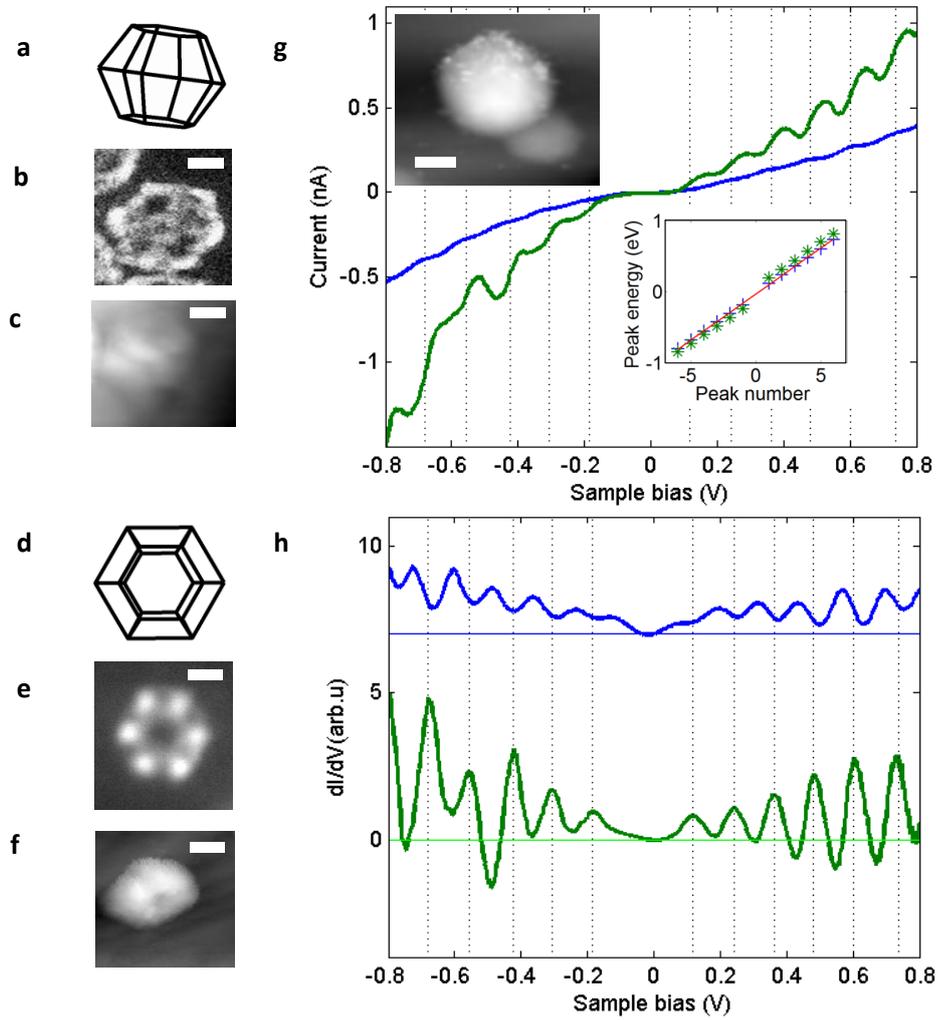

FIG 1. STS measurements on an empty Ru cage revealing negative tunneling conductance. (a-f) illustrations, STEM and STM images portraying two different projections of the Ru cage, one is emphasizing the median arm (a-c) and the other the pore of the cage (d-f). Scale bars for the STEM (b,e) and STM (c,f, g inset) images are 5nm. The gray scale range in the STM images is 0-5nm. (g) I-V curves and (h) corresponding dI/dV-V tunneling spectra, offset vertically for clarity, measured at 4.2 K on the same Ru cage, portraying Coulomb staircase (blue curves) and the NDC effect (green curves). Evidently, the staircase charging peaks correlate well with the NDC features. The green and blue spectra were measured at the same bias and current set-points, $V_s$ = 0.172V and $I_s$ = 49.6pA. The lower inset of (g) presents the peak bias values (for both sets) as a function of peak number (negative for negative bias values).

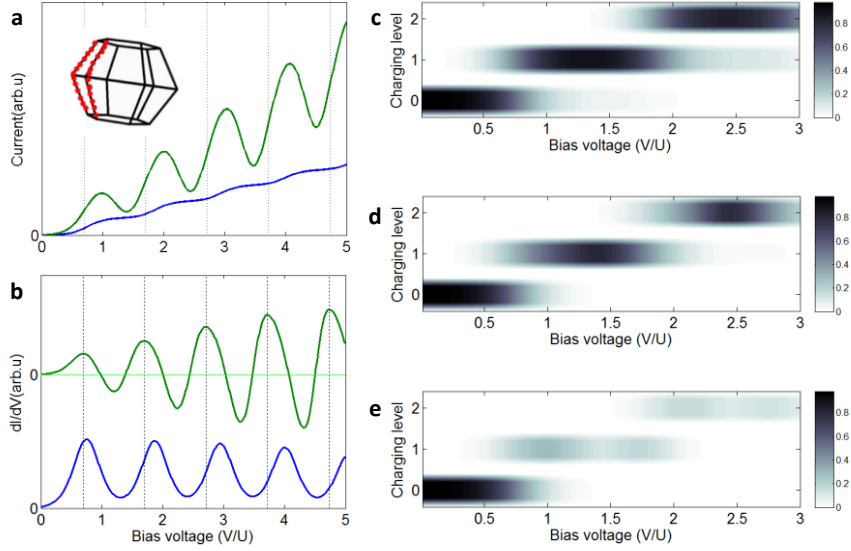

FIG 2. (a) Simulated I-V curves and (b) corresponding dI/dV-V curves calculated using a master equation approach as described in the text. The sample bias, V, is normalized to the single electron charging energy, U. The inset of (a) depicts the Ru-cage geometry, where the two coupled active conduction channels are drawn in red. In the case where the two coupled channels are both connected to the substrate and STM tip with similar tunneling resistances we observe a periodic Coulomb staircase with spacing $\sim U$ (blue curves). When one of the channels is effectively disconnected from either the STM tip or the substrate periodic NDC emerges (green curves). (c) presents the charging level occupation diagram for the case where both channels are equally coupled to the tip and the substrate and 'conventional' SET characteristic are observed. (d-e) correspond to the case where NDC is observed; (d) showing the level occupation of the disconnected cannel and (e) of the conducting one.

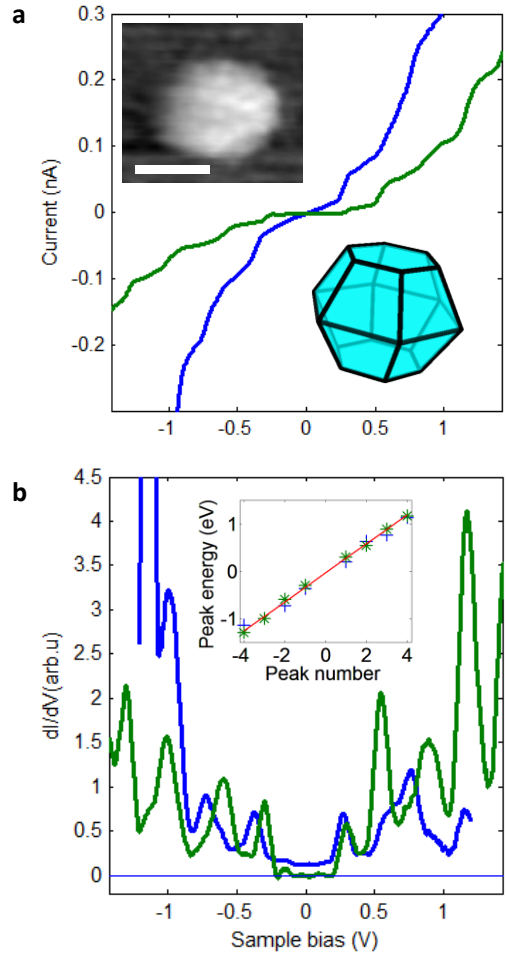

FIG 3. (a) I-V and (b) corresponding dI/dV-V tunneling spectra measured at 4.2 K on the Ru cage encapsulating the CU$_2$S core of a hybrid Cu$_2$S/Ru-cage QD (that is illustrated in the inset of a). The spectra manifest SET effects typical for metallic dots, the Coulomb blockade and staircase. The green and blue spectra were measured with different bias and current set-point (V$_s$=0.94V and 0.74 V and I$_s$=65pA and 78 pA, respectively), thus affecting the tip-cage distance and consequently the value of the 'residual offset charge', suppressing the blockade (SI Fig. S10). The periodicity of the staircase was only slightly affected. (a) top inset: depicts an STM images portraying a Cu$_2$S/Ru-cage. Scale bar for the STM image is 5nm and the gray scale range is 0-5nm. The inset of (b) presents the conductance-peak bias values plotted as a function of peak number (negative for negative bias values) for the two curves, in corresponding colors. The voltage differences between adjacent peaks are presented in Fig. S12.

**Periodic negative differential conductance in a single metallic nano-cage**

Yehonadav Bekenstein[1,2], Kathy Vinokurov[2], Tal J. Levy[3], Eran Rabani[3*], Uri Banin[2*], Oded Millo[1*]

[1]*Racah Institute of Physics and the Center for Nanoscience and Nanotechnology, The Hebrew University of Jerusalem, Jerusalem 91904, Israel*

[2] *Institutes of Chemistry and the Center for Nanoscience and Nanotechnology, The Hebrew University of Jerusalem, Jerusalem 91904, Israel*

[3] *School of Chemistry, Sackler Faculty of Exact Sciences, Tel Aviv University, Tel Aviv 69978, Israel*

*To whom correspondence should be addressed. e-mail: milode@vms.huji.ac.il , uri.banin@huji.ac.il, rabani@tau.ac.il

# Supporting information

## I. STM imaging of a Ru-cage QD deposited on Au(111) surface

For the STM measurements the QDs solutions were drop cast onto a flame annealed Au(111) substrate and let dry, after which the samples were promptly inserted into a homemade STM. The STM measurements were performed using Pt-Ir tips, in clean He exchange gas inserted into the sample space after evacuation. The Au(111) substrates exhibited atomic-scale roughness (steps and dislocations) even after QD deposition, indicating suitability for QD measurements (see Fig. S1).

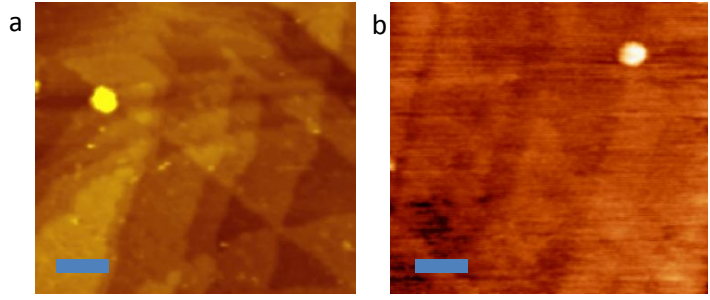

FIG S1. (a)(b) STM topographic image 150x150 nm$^2$ in size of a flame annealed Au(111) substrate on top of which (a) a single Ru cage is observed (marked by the arrow) and (b) a $Cu_2S$/Ru hybrid QD is observed . In Both micrographs atomic steps and surface dislocations (~2Å height) forming the triangular structure typical to the Au(111) surface are clearly seen, indicating the atomic-scale roughness of the substrate required for QD measurements. Scale bar is 20 nm.

## II. Model calculation details

The model consists of two channels representing two possible different pathways for transport provided by the Ru cage [see Fig. S2(a)]. Each channel is modeled by $N$ one-electron degenerate levels with a coupling between the levels of magnitude $U$ (the charging energy). In addition, the two channels are also coupled to each other by an interaction of magnitude $U_{int}$ (the interaction energy between channels, see also a discussion at the end of this section). A sketch describing the energetics of the model is shown in Fig. S2(b). In this model the STM tip and substrate are modeled as infinite non-interacting Fermionic baths [1-3]. The Hamiltonian in second quantization is given by

$$H = H_{cage} + H_{tip/sub} + H_{tunnel}, \qquad (1)$$

where

$$H_{cage} = \varepsilon \sum_{\nu=\alpha,\beta} \sum_i n_i^\nu + U \sum_{\nu=\alpha,\beta} \sum_{i>j} n_i^\nu n_j^\nu + U_{int} \sum_{i,j} n_i^\alpha n_j^\beta \qquad (2)$$

represents the electronic structure of the Ru cage. The first term in $H_{cage}$ stands for the non-interacting energy ($\varepsilon$) for each state $i$ on the two channels, $\nu = \alpha, \beta$, which also models the effects of the Fermi level offset affecting the Coulomb blockade. The parameter ($\varepsilon$) is related to the 'effective residual fractional charge', $Q_0$ discussed in section III below. The second term represents the sum of charging energies ($U$) of the two individual channels, where $n_i^\nu = 1$ or $0$ if state $i$ on channel $\nu$ is occupied by an electron or not, respectively. This term provides the Coulomb staircase in the case of uncoupled channels (physically and/or electrostatically). The third term represents the interaction energy ($U_{int}$) between electrons on the different channels of the cage.

The tip(substrate) electronic structure is represented as non-interacting Fermionic baths by

$$H_{tip/sub} = \sum_{k \in tip(sub)} \varepsilon_k \cdot n_k, \tag{3}$$

where, again, $n_k = 1$ or $0$ if state $k$ on the tip(substrate) is occupied by an electron or not, and $\varepsilon_k$ is the level energy. The last term in the Hamiltonian is the tunneling Hamiltonian [3],

$$H_{tunnel} = \sum_{\nu = \alpha, \beta} \left( \sum_{\substack{i \in \nu, \\ k \in tip(sub)}} t_{i,k}^\nu d_{i,\nu}^\dagger c_k + h.c. \right). \tag{4}$$

Here the operator $d_{i,\nu}^\dagger (d_{i,\nu})$ is the creation (annihilation) operator of an electron on channel $\nu$ state $i$, and $c_k (c_k^\dagger)$ is the annihilation (creation) operator of an electron on the tip/substrate in state $k$. The term $t_{i,k}^\nu$ denotes the coupling strength between the channels and the tip/substrate, and we generally used asymmetric couplings (i.e. $t_{i,sub}^\nu > t_{i,tip}^\nu$) to simulate more realistically the STM experiment. The bias voltage across the system (the Ru cage) is obtained by applying different Fermi levels to the tip and the substrate. We denote the Fermi levels by $\mu_{tip(sub)}$. To obtain the steady state current through the system under a given voltage bias we apply the master-equation formalism (ME) [4-6]. In the ME approach the steady state current can be calculated via

$$I_{tip} = \sum_{\tau, \zeta} S \cdot R_{tip}(\tau \to \zeta) \cdot P_\tau, \tag{5}$$

where $R_{tip}$ represents the part of the total transition rate $R$ associated with the STM tip, the indices $\tau$ and $\zeta$ stand for the multielectron states, $P_\tau$ is the probability of being in state $\tau$, and $S = -1$ or $1$ if state $\zeta$ has one electron less or more than state $\tau$, respectively. For the rates we write

$$R_{tip}(\tau \to \zeta) = \begin{cases} \frac{\gamma_{tip}}{\hbar} f_{\tau,\zeta}^{tip}(\epsilon_{\zeta\tau} - \mu_{tip}) & \text{if } \zeta \text{ has one electron more than } \tau, \\ \frac{\gamma_{tip}}{\hbar}\left(1 - f_{\tau,\zeta}^{tip}(\epsilon_{\tau\zeta} - \mu_{tip})\right) & \text{if } \zeta \text{ has one electron less than } \tau, \end{cases}$$

where $\epsilon_\tau$ is the energy of the multielectron state $\tau$, $\epsilon_{\tau\zeta} = \epsilon_\tau - \epsilon_\zeta$. $f_{\tau,\zeta}^{tip(sub)}(\epsilon_{\zeta\tau} - \mu_{tip(sub)})$ is the Fermi-Dirac distribution and $\frac{\gamma_{tip(sub)}}{\hbar} = \frac{2\pi}{\hbar}\sum_{k \in tip(sub)}|t_{i,k}^\nu|^2 \delta\left(\varepsilon - \varepsilon_k^{tip(sub)}\right)$ can be interpreted as the rate constant at which an electron escapes in and out of the system. We can calculate the individual probabilities by noting that under steady state conditions there must be no net flow into or out of any state. The probabilities are found by solving

$$\sum_\zeta R(\tau \to \zeta) P_\tau = \sum_\zeta R(\zeta \to \tau) P_\zeta , \tag{6}$$

where $R(\tau \to \zeta) = R_{tip}(\tau \to \zeta) + R_{sub}(\tau \to \zeta)$.

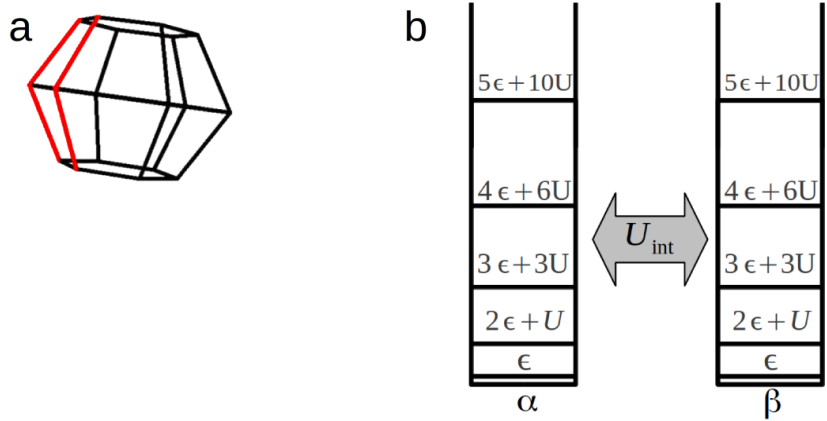

FIG S2. (a) A sketch of the Ru-cage geometry portraying two active conduction channels (in red). (b) Multi-electron level diagram of the two channels described in our model denoted $\boldsymbol{\alpha}$ and $\boldsymbol{\beta}$. The two channels are coupled by an inter-channel interaction (Coulomb repulsion) $\boldsymbol{U_{int}}$.

First, we consider the case where the two coupled channels are both connected to the substrate and STM tip with similar tunneling resistances. In Fig. S3(a) we plot the current versus the bias voltage for several different inter-channel interaction ($U_{int}$) values. Changing the value of $U_{int}$ mainly affects the magnitude of the current and the Coulomb staircase oscillation period. However, for this geometry, when both channels are connected to the tip and substrate, NDC is not observed irrespective of the mutual effective gating $U_{int}$. We also note that $U_{int}$, being an internal property of the cage, is not expected to change when one of the channels become disconnected, and thus the NDC and staircase periods should be nearly the same, as indeed observed in the experiment.

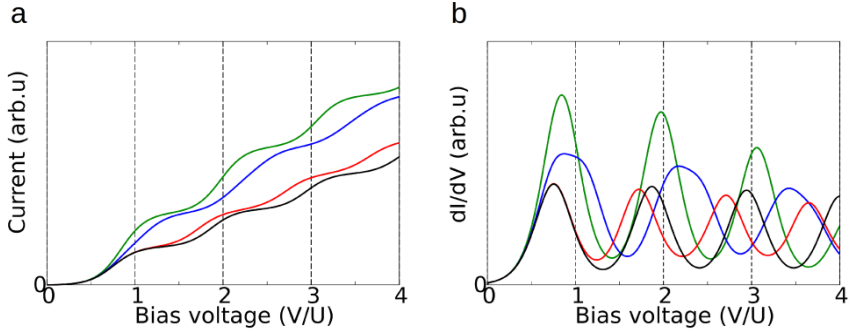

FIG S3. (a) I-V and b, dI/dV-V plots obtained from the ME approach for several different inter-channel interaction ($U_{int}$) values. Parameters used for all simulations (in units of U) are: $\gamma_{sub(tip)} = 0.1U$ ($0.05U$), and $\beta = \frac{1}{K_B T} = 7U^{-1}$. The green line is obtained with $U_{int} = 0$, the blue line with $U_{int} = 0.2U$, the red line with $U_{int} = 0.8U$, and the black line with $U_{int} = U$. As can be clearly seen, NDC is not observed in the case when both channels are effectively connected to the STM tip and substrate.

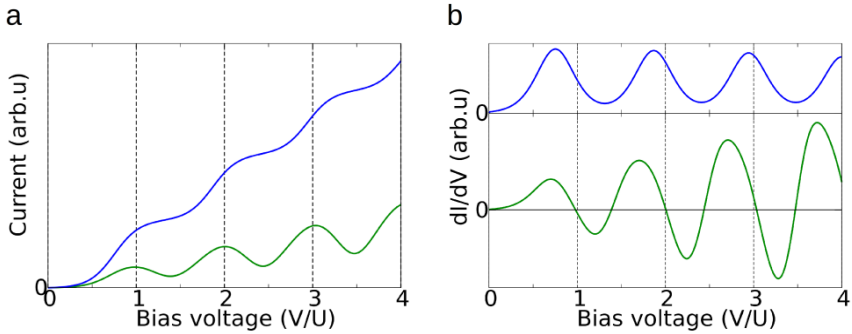

FIG S4. (**a**) I-V and (b) dI/dV-V plots obtained from the ME approach. Parameters used (in units of U) are: $\beta = 7U^{-1}$, $U_{int} = U$, and $\varepsilon = U$. The blue lines represent the Coulomb staircase situation obtained with $\gamma_{sub} = 0.1U$, and $\gamma_{tip} = 0.05U$ (for both channels). The green lines stand for the NDC situation where for the conducting channel we take $\gamma_{sub} = 0.1U$, and $\gamma_{tip} = 0.05U$ while for the blocked channel $\gamma_{sub} = 0.1U$, and $\gamma_{tip} = 0$.

Now consider the case where one of the channels is effectively disconnected from either the STM tip or the substrate (but not from both). This gives rise to periodic NDC, as can be seen in Fig. S4. For fixed parameters, the current in the NDC is smaller than the normal staircase. However, if transition from the staircase situation (where both channels are connected to the tip and substrate) to the NDC case (where one channel is effectively disconnected from either the tip or substrate) is accompanied by a decrease in the tunneling resistance then the situation is reversed as can be seen in Fig. S5. Such a scenario can be encountered in the experiment, e.g., by movement of the tip or reorientation of the Ru nano-cage. Another scenario encountered in the experiment is the change in the 'effective residual fractional charge', $Q_0$. This causes a phase shift of the peaks (towards both higher and lower bias voltage values). This effect is modeled here by changing the value of $\varepsilon$ and presented below in section VI, along with corresponding experimental spectra.

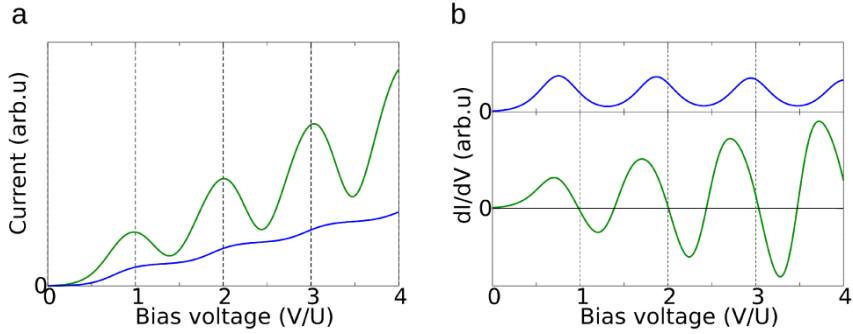

FIG S5. (a) I-V and (b) dI/dV-V plots obtained from the ME approach. Parameters used (in units of U) are: $\boldsymbol{\beta} = 7U^{-1}$, $\boldsymbol{U_{int}} = \boldsymbol{U}$, and $\boldsymbol{\varepsilon} = \boldsymbol{U}$. Here we assumed that the transition from the Coulomb staircase situation ($\boldsymbol{\gamma_{sub}} = \boldsymbol{0.1U}$, and $\boldsymbol{\gamma_{tip}} = \boldsymbol{0.05U}$ for both channels) is accompanied by a decrease in the tunneling resistance thus changing the coupling to the STM for the NDC situation. For this simulation we chose $\boldsymbol{\gamma_{sub}} = \boldsymbol{0.8U}$, and $\boldsymbol{\gamma_{tip}} = \boldsymbol{0.4U}$ for the conducting channel while for the blocked channel $\boldsymbol{\gamma_{sub}} = \boldsymbol{0.8U}$, and $\boldsymbol{\gamma_{tip}} = \boldsymbol{0}$.

Disconnecting one of the channels from either the STM tip or the substrate by itself is not sufficient for the appearance of periodic NDC (or even just NDC). This is easily understood when considering the extreme scenario where $U_{\text{int}} = 0$. In this event the conducting channel is not affected by the blocked one, resulting in a periodic Coulomb staircase. In Fig. S6 we show the dependence of NDC on the magnitude of the inter-channel interaction, $U_{\text{int}}$, for the case where one of the channels is effectively disconnected. We find that NDC disappears all together for a value of $U_{\text{int}}$ smaller than approximately $0.6U$.

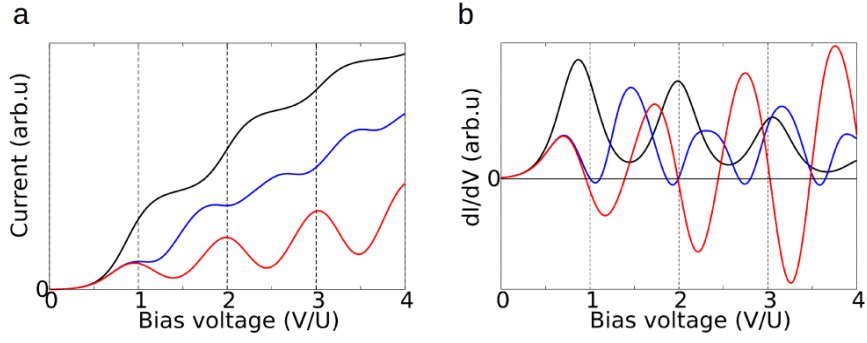

FIG S6. (a) I-V and (b) dI/dV-V plots obtained from the ME approach for different values of the inter-channel interaction $U_{int}$ for the case where one of the channels is effectively disconnected. Parameters used for all simulations (in units of U) are: $\gamma_{sub(tip)} = 0.1U$ ($0.05U$) for the conducting channel, $\gamma_{sub(tip)} = 0.1U$ ($0U$) for the blocked channel and $\beta = 7U^{-1}$. The black line is obtained with $U_{int} = 0$, the blue line with $U_{int} = 0.6U$, and the red line with $U_{int} = U$. Periodic NDC emerges for values of $U_{int}$ bigger than ~$0.6U$, while it completely disappears for $U_{int} < 0.6U$.

The model can be easily expanded to include more than two channels (three and four channels). In the extended model we add one more parameter $U_{int}^{scr}$ (screened inter-channel interaction), which represents the magnitude of the inter-channel interaction between non-adjacent channels while $U_{int}$ remains the interaction parameter between adjacent channels, and we note $U_{int}^{scr} < U_{int}$. This underlines the assumption that there is a screening effect between distant channels. While no NDC is observed when all channels are effectively connected (consistent with the two channels model), periodic NDC appears as soon as at least one channel is disconnected and the inter-channel interactions are not too weak as already discussed above. In Fig. S7 we plot the I-V characteristics for the three and four channels models.

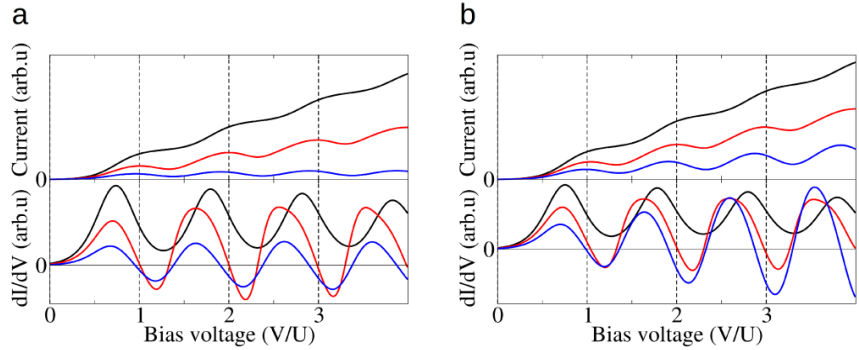

FIG S7. I-V curves for the case of three and four channels. While no NDC appears when all channels are connected, periodic NDC emerges as soon as at least one of the channels is effectively disconnected. (a,b) I-V (upper inset) and dI/dV-V (lower inset) plots obtained from our expanded model. In (a) we show the results calculated from the three channels model. The black line corresponds to the case where all channels are connected; red line corresponds to the case where one channel is disconnected, while the blue line corresponds

to the case where two out of the three channels are disconnected. (b) Results calculated from our four channels model. The black line corresponds to the case where all channels are connected; red line corresponds to the case where one channel is disconnected, while the blue line corresponds to the case where two out of the four channels are disconnected. Parameters used for all simulations are: $\beta = 7U^{-1}$, $U_{int} = U$, $U_{int}^{scr} = 0.85U$, $\gamma_{sub(tip)} = 0.1U$ ($0.05U$) for the conducting channels and $\gamma_{sub(tip)} = 0.1U$ ($0U$) for the blocked channels.

## III. A classical electrostatic model used to estimate $U_{int}$

We argue that in the vicinity of junctions between arms of the Ru cages, where the distance between two neighboring arms in the nano-cage is small, $U_{int}$ is comparable to $U$, and thus the condition $U_{int} > 0.6U$ is indeed soundly satisfied. To show this, we model (two) paths (arms of the cage) as (two) finite cylinders of radius $R$, length $L$, surface charge density $\sigma$ placed a distance $d$ apart (Fig. S8):

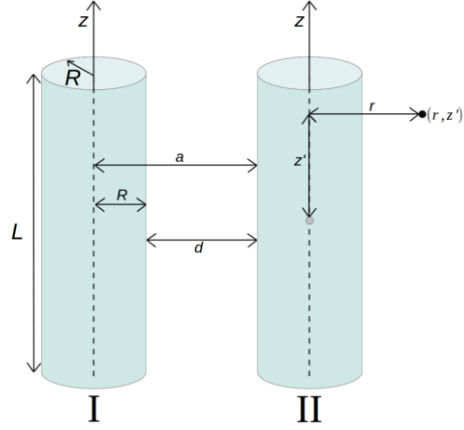

FIG S8. Our model constitute two conducting finite cylinders (labeled I and II) of radius $R$, length $L$, surface charge density $\sigma$, that are placed a distance $d$ apart. The cylinders model arms of the nano-cage.

The potential of such a cylinder is given by

$$\Phi(r,z) = \int_0^{2\pi} d\phi_1 \int_{-L/2}^{L/2} dz_1 \frac{\sigma R}{\sqrt{R^2 + r^2 - 2 \cdot R \cdot r \cdot cos(\phi_1) + (z_1 - z)^2}}, \quad r \geq R$$

The work required to bring a charge $q$ from "infinity" to the surface of cylinder I, for the given configuration depict in figure S8, is

$$W = \Phi_I(R, z) \cdot q + \Phi_{II}(a, z) \cdot q.$$

The first term corresponds to the charging energy $U$ while the second term to the interaction energy $U_{int}$ (i.e., the repulsion caused by cylinder II). We wish to calculate the distance $d$ ($d = a - R$) below which

$$U_{int} > 0.6U \rightarrow \frac{U_{int}}{U} > 0.6, \text{ meaning}$$

$$\frac{U_{int}}{U} = \frac{\Phi_{II}(a,z)}{\Phi_{I}(R,z)} \equiv F(d,z) > 0.6$$

In Fig. S9 we plot $F(d,z)$ ($\frac{U_{int}}{U}$) as a function of $d$ for different values of $z$. We note in passing that the integral were solved numerically. It is clearly seen that $U_{int} > 0.6U$ for distances $d$ smaller than $5R$. Our nano-cages always satisfy the latter condition ($d < 5R$), in particular in the vicinity of a junction where neighboring arms meet.

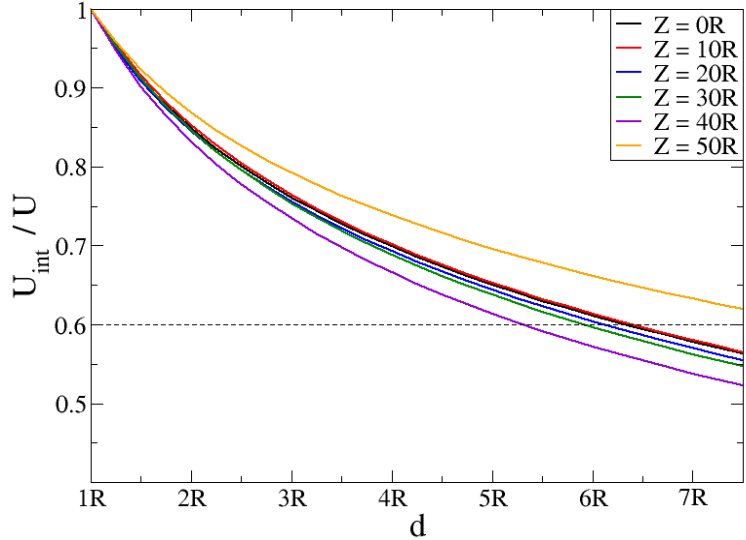

FIG S9. Plots of $(d,z)$, or $\frac{U_{int}}{U}$, as a function of $d$ for different values of $z$. For these calculations we chose $L = 100R$. In general we found that the longer the cylinder is, the further apart the cylinders can be placed while still fulfilling the condition $\frac{U_{int}}{U} > 0.6$.

## IV. SET effect measured on a hybrid Cu₂S/Ru-cage QD and fits to the 'orthodox model'

The tunneling spectra measured on the hybrid QDs varied significantly for different tip locations along the dot, as will be detailed elsewhere. When the tip was positioned above the encapsulating Ru cage, 'conventional' SET effects, the Coulomb blockade and staircase, were observed. In order to verify that the gap around zero bias is associated with the Coulomb blockade (and not to the semiconducting Cu₂S gap), we followed the procedure described in Ref. 7. By varying the tip-QD distance and thus the capacitance value corresponding to this junction, we can control the 'effective residual charge', $Q_0$, on the dot, an effect similar to changing the gate in a single electron tunneling transistor. $Q_0$ is given by: $Q_0 = (C_1 \Delta \Phi_1 - C_2 \Delta \Phi_2)/e$, where $C_i, \Delta \Phi_i, i = 1,2$, are the capacitances and contact potentials associated with the substrate-QD and QD-tip junctions, respectively. When mod($Q_0$) = 0.5$e$, the Coulomb blockade is suppressed and finite conductance is measured at zero bias Fig. S10(a).

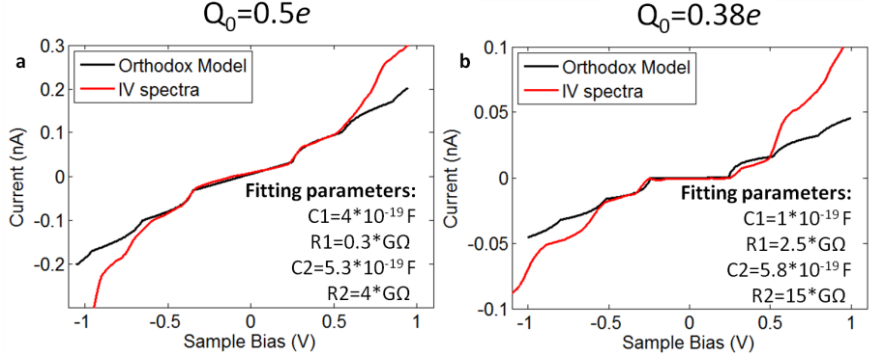

FIG S10. (a,b) Tunneling I-V characteristics measured at 4.2 K on a hybrid Cu$_2$S/Ru QD, the same as those presented in Figure 3 in the paper (red curves) fitted to the "orthodox model" of SET (black curves). Bests fit to the I-V curve presented in (a) (measured with a setting of $V_s$ = 0.73 V and $I_s$ = 78 pA), for which the Coulomb blockade is suppressed, resulting in finite zero-bias conductance, was obtained for $Q_0$ = 0.5$e$. In contrast, a pronounced Coulomb blockade is found for the spectrum shown in (b) (measured with $V_s$ = 0.94 V and $I_s$ = 65 pA), and the fit yields $Q_0$ = 0.38$e$. The capacitance and resistance values of the DBTJ junction parameters are given in the panels. The increased conductance of the experimental data with respect to the fit at bias voltages larger than 0.5 V is due to the reduction in the tunneling barrier (increase of the tunneling matrix element), which was not taken into account in the fit. The theoretical curves were calculated using the formalism detailed in Ref. [7].

## V. SET effect measured on Ru-cage QD fitted to the 'Orthodox Model'

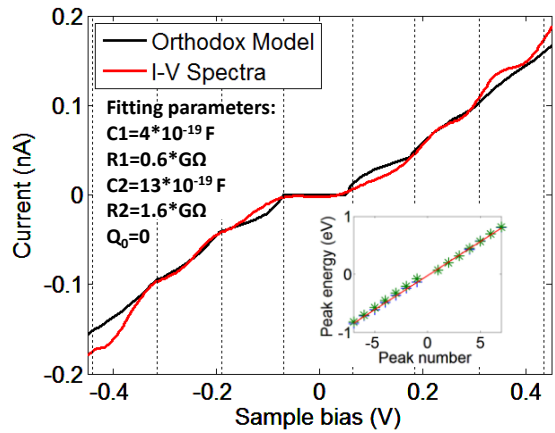

FIG S11. Tunneling I-V characteristics measured at 4.2 K on a Ru cage QD (red curve), the same data as presented in Figure 1 in the paper (blue curve) fitted to the "orthodox model" of SET (black curves). Best fit to the I-V curve (measured with a setting of $V_s$ = 0.172 V and $I_s$ = 49.6 pA), was obtained for $Q_0$ = 0. The capacitance and resistance values of the DBTJ junction parameters are given in the panel. The increased conductance of the experimental data with respect to the fit at bias voltages larger than 0.5 V is due to the reduction in the tunneling barrier (increase of the tunneling matrix element), which was not taken into account in the fit. The theoretical curves were calculated using the formalism detailed in Ref. [7].

## VI. Statistics of charging energies and QD size distribution for both empty Ru cages and hybrid $Cu_2S$-Ru hybrid QDS

In this section we present the size distributions of the Ru nano-cages and $Cu_2S$/Ru hybrid QDs [Fig. S12(a)], along with the distribution of charging energy measured on 20 empty Ru nano-cages and 40 $Cu_2S$/Ru hybrid QDs [Fig. S12(a)]. It is quite evident that the rather narrow size distribution cannot account for the large spread in charging energies.

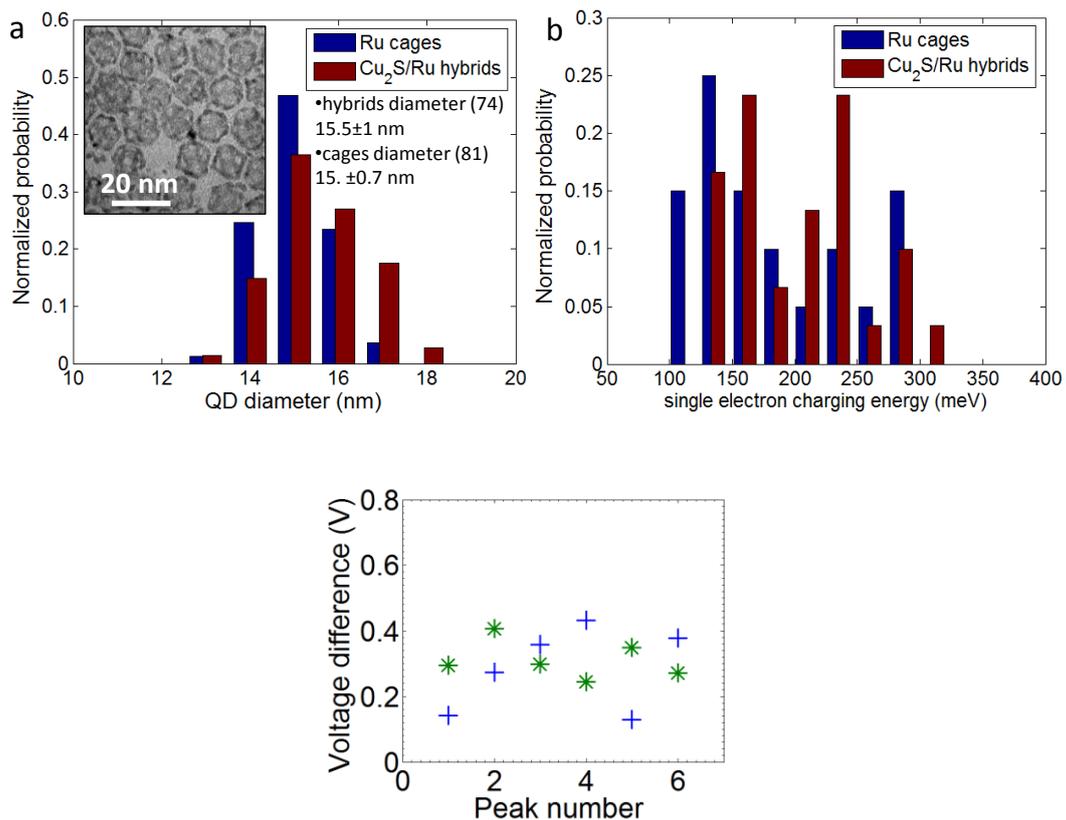

FIG S12. (a) Histograms of the diameters of the Ru nano-cages (represented by blue bars) and $Cu_2S$/Ru QDs (red bars). inset: a TEM micrograph depicting the monodispersesity in size of the Ru cages. (b) Histograms of the single electron charging energies deduced from the tunneling spectra, as explained in the paper. 20 Ru cages (blue

bars) and 40 Cu$_2$S/Ru hybrids (red bars) were measured. (c) Voltage differences between adjacent peaks in the tunneling spectra of a Cu$_2$S/Ru hybrid presented Fig. 3 of the paper, showing that the charging energies are very similar in the two settings.

## VII. Additional NDC results demonstrating different peak shifts

As discussed in the paper, the NDC peaks usually slightly shift with respect to those of the 'conventional' SET dI/dV-V tunneling spectra. These shifts most probably result from the changes in the junction parameters (which affect, in turn, Q$_0$ and the charging energy) that take place during the transition from the SET to NDC behaviors. It is therefore expected that also the shifts will vary from on measurements to another, as indeed becomes evident by comparing between the two additional experimental and theoretical tunneling spectra presented in Fig. S13(a,b) and between them and those presented in the paper. The different theoretical curves presented in Fig. S13(c) were calculated with different values of $\varepsilon$ giving rise to different shifts. In particular, the NDC peaks can either advance or lag behind the SET peaks.

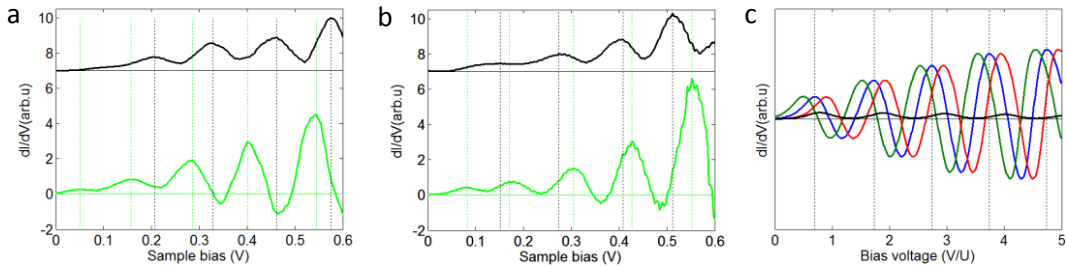

FIG S13. Experimental dI/dV-V spectra (a,b) measured on two different Ru nano-cages, showing different shifts between the 'conventional' SET (black curves) and NDC (green curves) peaks, where in one (a) the NDC peaks lag behind those of the SET (b) and vice-versa in the second. (c) Calculated spectra showing 'conventional' SET (black curve) and NDC behaviors with different peak shifts, resulting from small changes in the values of $\varepsilon$. The black and blue curves are calculated using $\boldsymbol{\varepsilon = U}$, green curve is obtained using $\boldsymbol{\varepsilon = 0.8U}$, and the red curve is calculated using $\boldsymbol{\varepsilon = 1.2U}$